\documentclass[aps,pra,superscriptaddress,amsmath,amssymb,preprintnumbers
,showkeys]{revtex4}
\usepackage{amssymb}
\usepackage{amsmath}
\usepackage{bm}
\usepackage{color}
\usepackage{graphicx}

\def\Ek{{\cal E}_\kappa}
\def\Ekp{{\cal E}_{\kappa'}}

\begin{document}


\title{Completeness of scattering states of the Dirac Hamiltonian with a step potential}


\author{M. Ochiai}
\author{H. Nakazato}
\affiliation{Department of Physics, Waseda University, Tokyo 169-8555, Japan}


\begin{abstract}
The completeness, together with the orthonormality, of the eigenfunctions of  the Dirac Hamiltonian with a step potential is shown explicitly.
These eigenfunctions describe the scattering process of a relativistic fermion off the step potential and the resolution of the identity in terms of them (completeness) is shown by explicitly summing them up, where appropriate treatments of the momentum integrations are crucial.
The result would bring about a basis on which a field theoretical treatment for such a system can be developed.
\end{abstract}

\keywords{completeness, Dirac Hamiltonian, step potential}

\maketitle


\section{Introduction}
In quantum mechanics, a physical observable is represented by a self-adjoint operator acting on a Hilbert space.
Its eigenvalues are all real-valued, corresponding to its measured values in experiments, and the eigenfunctions belonging to different eigenvalues are orthogonal to each other, which is in accord with the physical situation that the two events with different outcomes are mutually exclusive and never happen simultaneously.
These statements are easily proved and the proofs are found in any textbook of quantum mechanics, in which, however, we always find another statement accompanied, claiming that the eigenfunctions form a complete set.
The statement is physically sound because if we measure a physical quantity in any state, we would and should have a result, the obtained value of which is one of the eigenvalues of the corresponding operator and this implies that any state can be expanded in terms of the eigenfunctions. 
Even though we would all admit its necessity for a physical ground, the proof of completeness of eigenfunctions would not so easily be done once they constitute a continuous spectrum.
This constitutes a striking contrast to the case where only a finite number of discrete states exist, for in such a case one can convince oneself that a finite number of mutually independent vectors are enough to express any state, thus the proof of their completeness is trivial.

When a continuous spectrum is involved, only exception where one can easily discuss and confirm the completeness of the eigenfunctions would be the free case.
In this case, the plane waves $\phi_{\bm k}({\bm x})$, i.e., the eigenfunctions of the free Hamiltonian $H={\bm p}^2/2m$ form a complete orthonormal set in the sense ($\hbar=1$) 
 \begin{equation}
 \int d^3{\bm k}\phi_{\bm k}({\bm x})\phi_{\bm k}^*({\bm x'})=\delta^3({\bm x}-{\bm x'}),\quad
\int d^3{\bm x}\phi_{\bm k}^*({\bm x})\phi_{\bm k'}({\bm x})=\delta^3({\bm k}-{\bm k'}),\quad
\phi_{\bm k}({\bm x})={1\over\sqrt{(2\pi)^3}}e^{i{\bm k}\cdot{\bm x}},
\label{eq:pw}
\end{equation}
which are essentially the (inverse) Fourier transforms of Dirac's delta function, or the resolution of the identity.
What is not trivial is to show these relations when the plane wave solutions are replaced with the scattering wave functions for a stationary potential problem.
As a matter of fact, even though the asymptotic completeness of the scattering states is proved within the framework of  ``formal scattering theory" (see, e.g., \cite{ReedSimon1979}) and there exists a classic review on the analytical properties of radial wave functions by Newton \cite{Newton1960}, the completeness of, say, the Coulomb scattering wave functions is shown on the basis of Newton's method rather recently \cite{MukhamedzhanovAkin2008}.
This is because the long-range potentials were not covered by their treatments and to show it on the basis of Newton's method, one has to evaluate rather involved integrals in the complex $k$-plane.
Surprisingly enough, the situation is not at all better for cases with simple potentials, say, the step or well potentials in one dimension. 
In fact, the proof of the completeness of one-dimensional scattering wave functions off a step potential appeared only in the late 80's \cite{TrottTrottSchnittler1989} and a mathematically more rigorous treatment for step and well potentials has to be waited until 2010 \cite{PalmaPradoReyes2010}.
Apparently these potentials do not satisfy some of the premises assumed for the general proofs and one has to work out such cases separately even though the potentials are very simple and considered rather elementary. 

In this paper, we shall demonstrate that the scattering wave functions of a Dirac Hamiltonian for a system with a step potential form a complete orthonormal set, along the same line of thought as \cite{TrottTrottSchnittler1989}, by summing up all of them.
Though the completeness of them in the same system is already shown in \cite{RuijsenaarsBongaarts1977} in a mathematically rigorous way, it is still of interest and instructive to see explicitly that the eigenfunctions of an interacting Dirac Hamiltonian can replace the plane waves in (\ref{eq:pw}) to play the same role.
The procedure taken here is intuitive, straightforward, though much involved and in accord with what physicists would expect.

The paper is organized as follows.
After the introduction of basic ingredients for the free Dirac Hamiltonian in the next section \ref{sec:DiracH}, the scattering wave functions of the Dirac Hamiltonian for the system with a finite step potential, responsible for an impulsive force at the origin, are presented in Sec.~\ref{sec:ScatteringStates}.
They are categorized into several cases depending on the boundary conditions, i.e., if a Dirac fermion with a positive (or negative) frequency is incident from the left or the right of the step potential, and on the value of incident energy.
The completeness condition expressed in the coordinate space refers to two spatial points and since the potential takes different values at the left and the right of the potential, it is examined for three different cases in Sec.~\ref{subsec:proof}.
The orthogonality between the left-incident and right-incident wave functions is shown explicitly for a particular case as an illustration in Sec.~\ref{subsec:orthogonal}.
The last section \ref{sec:summary} is devoted to a summary and prospect.

\section{Solutions of Free Dirac Equation}
\label{sec:DiracH}

For later convenience, we first fix the notations and normalizations of four-component spinors of the free Dirac Hamiltonian.
The stationary plane wave solutions for the free Dirac Hamiltonian
\begin{equation}
H_0=-i\alpha_z\partial_z+\beta m,
\quad\alpha_z=\begin{pmatrix}0&\sigma_z\\\sigma_z&0\end{pmatrix},
\quad\beta=\begin{pmatrix}\openone&0\\0&-\openone\end{pmatrix}
\end{equation}
are
\begin{equation}
u(p,s)e^{-ip\cdot z}=\sqrt{E_p+m\over2m}\begin{pmatrix}\openone\\{\sigma_zp\over E_p+m}\end{pmatrix}\bm\xi(s)e^{ipz-iE_pt},\quad
H_0u(p,s)e^{-ip\cdot z}=E_pu(p,s)e^{-ip\cdot z},
\end{equation}
for a positive frequency $E_p=\sqrt{p^2+m^2}$ and
\begin{equation}
v(p,s)e^{ip\cdot z}=\sqrt{E_p+m\over2m}\begin{pmatrix}{\sigma_zp\over E_p+m}\\\openone\end{pmatrix}\bm\xi(s)e^{-ipz+iE_pt},\quad
H_0v(p,s)e^{ip\cdot z}=-E_pv(p,s)e^{ip\cdot z},
\end{equation}
for a negative frequency $-E_p$, where $\sigma_z$ is the third Pauli matrix, $\openone$ the $2\times2$ unit matrix and $\bm\xi(s)$ a two-component spinor.
Note that since we will consider a step potential, we setup the $z$-axis at right angle with the boundary of the potential  and the trivial dependence on the other coordinates $x$ and $y$ is suppressed from the beginning and is ignored completely.
The problem is essentially in one spatial dimension, though the spinors live in four-dimensional space-time and  have four components.
These spinors are normalized and form a complete orthonormal set
\begin{equation}
\bar u(p,s)u(p,s')=\delta_{s,s'},\quad \bar v(p,s)v(p,s')=-\delta_{s,s'},\quad
u^\dagger(p,s)u(p,s')=v^\dagger(p,s)v(p,s')={E_p\over m}\delta_{s,s'},
\label{eq:udu&vdv}
\end{equation}
\begin{equation}
\sum_s[u(p,s)\bar u(p,s)-v(p,s)\bar v(p,s)]=\openone_{4\times4}.
\label{eq:uubar-vvbar}
\end{equation}
Now, if a constant potential is added to the above Hamiltonian in all space 
\begin{equation}
H_0'=H_0+V_0=-i\alpha_z\partial_z+\beta m+V_0,\quad(-\infty<z<\infty),
\end{equation}
the corresponding plane wave solutions are ($E_q=\sqrt{q^2+m^2}$)
\begin{equation}
u(q,s)e^{-iq\cdot z}=\sqrt{E_q+m\over2m}\begin{pmatrix}\openone\\{\sigma_zq\over E_q+m}\end{pmatrix}\bm\xi(s)e^{iqz-i(E_q+V_0)t}
\end{equation}
for a ``positive" frequency $E_q+V_0\ge V_0$ and
\begin{equation}
v(q,s)e^{iq\cdot z}=\sqrt{E_q+m\over2m}\begin{pmatrix}{\sigma_zq\over E_q+m}\\\openone\end{pmatrix}\bm\xi(s)e^{-iqz-i(-E_q+V_0)t}
\end{equation}
for a ``negative'' (relative to $V_0$) frequency $-E_q+V_0\le V_0$.
Notice that the frequency $-E_q+V_0$ can be positive for small momenta $|q|<\sqrt{V_0^2-m^2}$, even though it is to be associated with the spinor $v$.
Note also that the above explicit forms of spinors are not unique since $\sigma_z$ is commutable with the above $H_0$ and $H_0'$ and therefore, e.g., $\sigma_z u(p,s)$ can be used instead of $u(p,s)$.

\section{Scattering states}
\label{sec:ScatteringStates}
A scattering problem of a Dirac fermion with mass $m$ off a step potential
\begin{equation}
V(z)=\theta(z)V_0,\quad V_0>2m,
\end{equation}
where $\theta(z)$ is the Heaviside step function, is easily solved.
(For simplicity and definiteness, only those cases where the potential barrier is higher than $2m$ are considered.)
The stationary solutions for the left- and right-incident scattering problems, described by the total Hamiltonian
\begin{equation}
H=H_0+V(z)=-i\alpha_z\partial_z+\beta m+\theta(z)V_0
\end{equation}
are given by the eigenstates of the Hamiltonian $H$ and are characterized by their boundary conditions, i.e., left-incident ($\psi$) or right-incident ($\phi$), and the (range of) value of their eigenvalue $E$.

\begin{figure}[h]
\begin{center}
\includegraphics[width=6in]{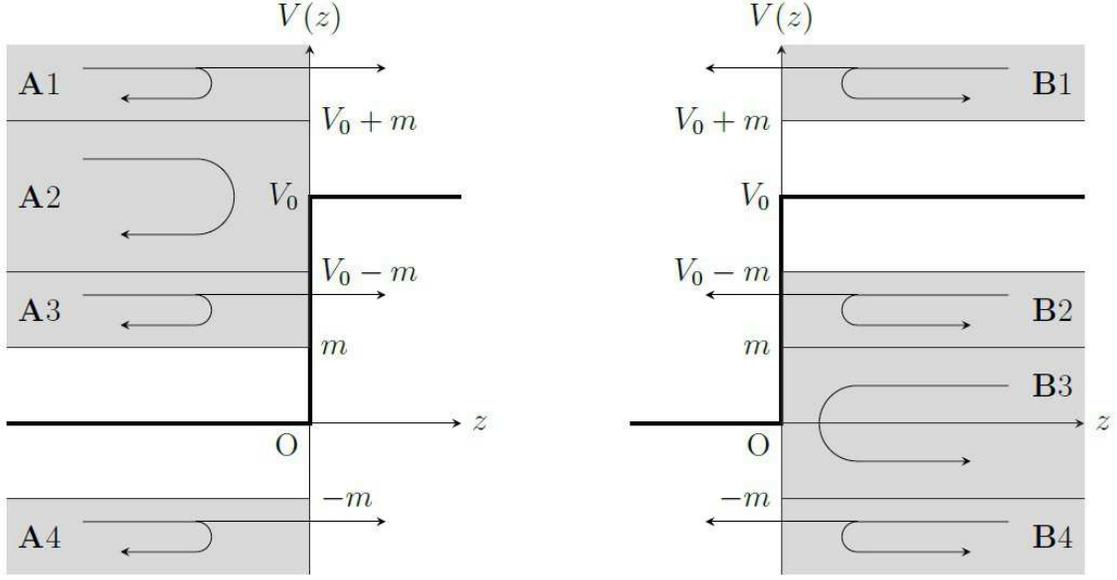}
\caption{Scattering states are categorized according to whether a particle is incident from the left ({\bf A}: $\psi$) or right ({\bf B}: $\phi$) of the step potential and to the ranges of their energies ({\bf1$\sim$4}).
No left(right)-incident scattering solution exists for the energy range $|E|<m\ (|E-V_0|<m)$.}
\end{center}
\end{figure}

\subsection{Left-incident cases $\psi^{(E)}(z,t)$}
\noindent
{\bf A1: for $\bm{E>V_0+m}$} ($E=E_p=E_q+V_0$ and therefore $p=\sqrt{E^2-m^2},\,q=\sqrt{(E-V_0)^2-m^2}$), the left-incident wave function reads as
\begin{equation}
\psi^{(E)}(z,t)={1\over\sqrt{2\pi}}\sqrt{m\over E}\left[\theta(-z)\bigl\{u(p,s)e^{ipz}+Ru(-p,s)e^{-ipz}\bigr\}+\theta(z)Tu(q,s)e^{iqz}\right]e^{-iEt}, 
\end{equation}
with the reflection and transmission coefficients 
\begin{equation}
R={\sqrt{E-V_0+m\over E+m}-\sqrt{E-V_0-m\over E-m}\over\sqrt{E-V_0+m\over E+m}+\sqrt{E-V_0-m\over E-m}}
,\quad
T={2\over\sqrt{E-V_0+m\over E+m}+\sqrt{E-V_0-m\over E-m}}
\label{eq:originalRT+}
\end{equation}
determined by the continuity condition at $z=0$.
Here and in the followings, the dependence on the spin $s$ is and will be suppressed in the wave functions.
The continuity of the current defined as $j_z=\bar\psi\gamma^3\psi=\psi^\dagger\alpha_z\psi$ at $z=0$ reads as
\begin{equation}
{p\over m}-|R|^2{p\over m}=|T|^2{q\over m},
\end{equation}
which expresses the conservation of the probability
\begin{equation}
P_r=|R|^2,\quad P_t={q\over p}|T|^2,\quad P_r+P_t=1.
\end{equation}
\noindent
{\bf A2: for \hbox{$\bm{V_0+m>E>V_0-m(>m)}$}}, no oscillating solution exists for the region $z>0$ and we have ($q=i\kappa'=i\sqrt{m^2-(E-V_0)^2}$)
\begin{equation}
\psi^{(E)}(z,t)={1\over\sqrt{2\pi}}\sqrt{m\over E}\left[\theta(-z)\bigl\{u(p,s)e^{ipz}+Ru(-p,s)e^{-ipz}\bigr\}+\theta(z)Tu(i\kappa',s)e^{-\kappa' z}\right]e^{-iEt}, 
\end{equation}
for $E>V_0$ and
\begin{equation}
\psi^{(E)}(z,t)={1\over\sqrt{2\pi}}\sqrt{m\over E}\left[\theta(-z)\bigl\{u(p,s)e^{ipz}+Ru(-p,s)e^{-ipz}\bigr\}+\theta(z)Ti\sigma_zv(-i\kappa',s)e^{-\kappa' z}\right]e^{-iEt}, 
\end{equation}
for $E<V_0$, with the reflection and transmission coefficients 
\begin{equation}
R={\sqrt{E-V_0+m\over E+m}-i\sqrt{m-(E-V_0)\over E-m}\over\sqrt{E-V_0+m\over E+m}+i\sqrt{m-(E-V_0)\over E-m}},\quad
T={2\over\sqrt{E-V_0+m\over E+m}+i\sqrt{m-(E-V_0)\over E-m}},
\end{equation}
resulting in a total reflection
\begin{equation}
P_r=|R|^2=1,\quad P_t=0,
\end{equation}
for the current in the region $z>0$ vanishes identically.

\noindent
{\bf A3: for $\bm{V_0-m>E>m}$}, the solution in the region $z>0$ is expressed in terms of $v$ ($E=E_p=-E_q+V_0$ or $q=\sqrt{(V_0-E)^2-m^2}$)
\begin{equation}
\psi^{(E)}(z,t)={1\over\sqrt{2\pi}}\sqrt{m\over E}\left[\theta(-z)\bigl\{u(p,s)e^{ipz}+Ru(-p,s)e^{-ipz}\bigr\}+\theta(z)T\sigma_zv(q,s)e^{-iqz}\right]e^{-iEt}, 
\label{eq:Klein}
\end{equation}
with
\begin{equation}
R={\sqrt{V_0-E-m\over E+m}-\sqrt{V_0-E+m\over E-m}\over\sqrt{V_0-E-m\over E+m}+\sqrt{V_0-E+m\over E-m}},\quad
T={2\over\sqrt{V_0-E-m\over E+m}+\sqrt{V_0-E+m\over E-m}}.
\end{equation}
The continuity of current implies the conservation of probability
\begin{equation}
{p\over m}-|R|^2{p\over m}=|T|^2{q\over m}\quad\to\quad P_r+P_t=|R|^2+{q\over p}|T|^2=1.
\label{eq:cccp}
\end{equation}
Recall that the ordinary conservation law of probability is satisfied, but a finite and non-vanishing transmission probability survives even at the infinite-potential limit $V_0\to\infty$
\begin{equation}
P_t= {q\over p}|T|^2\longrightarrow{2\sqrt{E^2-m^2}\over E+\sqrt{E^2-m^2}}\not=0.
\label{eq:KleinTunneling}
\end{equation}
This phenomenon is known as the Klein tunneling \cite{DombeyCalogeracos1999}.
Notice that no left-incident particle is allowed for $m>E>-m$.

\noindent
{\bf A4: for $\bm{-m>E}$}, the eigenstates are given by the negative-frequency solutions ($E=-|E|=-E_p=-E_q+V_0$)
\begin{equation}
\psi^{(E)}(z,t)={1\over\sqrt{2\pi}}\sqrt{m\over|E|}\left[\theta(-z)\bigl\{v(p,s)e^{-ipz}+Rv(-p,s)e^{ipz}\bigr\}+\theta(z)Tv(q,s)e^{-iqz}\right]e^{i|E|t}, 
\end{equation}
with
\begin{equation}
R={\sqrt{|E|+V_0+m\over|E|+m}-\sqrt{|E|+V_0-m\over|E|-m}\over\sqrt{|E|+V_0+m\over|E|+m}+\sqrt{|E|+V_0-m\over|E|-m}},\quad
T={2\over\sqrt{|E|+V_0+m\over|E|+m}+\sqrt{|E|+V_0-m\over|E|-m}}.
\label{eq:originalRT-}
\end{equation}
The current continuity and the conservation of probability are the same as in (\ref{eq:cccp}).

\subsection{Right-incident cases $\phi^{(E)}(z,t)$}

\noindent
{\bf B1: for $\bm{E>V_0+m}$}, $E=E_q+V_0=E_p$ (or $p=\sqrt{E^2-m^2},\,q=\sqrt{(E-V_0)^2-m^2}$) and the right-incident wave function is given by
\begin{equation}
\phi^{(E)}(z,t)={1\over\sqrt{2\pi}}\sqrt{m\over E-V_0}\left[\theta(-z)Tu(-p,s)e^{-ipz}+\theta(z)\bigl\{u(-q,s)e^{-iqz}+Ru(q,s)e^{iqz}\bigr\}\right]e^{-iEt}, 
\end{equation}
with  
\begin{equation}
R={\sqrt{E+m\over E-V_0+m}-\sqrt{E-m\over E-V_0-m}\over\sqrt{E+m\over E-V_0+m}+\sqrt{E-m\over E-V_0-m}},\quad
T={2\over\sqrt{E+m\over E-V_0+m}+\sqrt{E-m\over E-V_0-m}}.
\end{equation}
The continuity of current and the conservation of probability read as
\begin{equation}
-|T|^2{p\over m}=-{q\over m}+|R|^2{q\over m},\quad
P_r+P_t=|R|^2+{p\over q}|T|^2=1.
\label{eq:cccp2}
\end{equation}

\noindent
Observe that  no right-incident particle is allowed for $V_0+m>E>V_0-m$.

\noindent
{\bf B2: for $\bm{V_0-m>E>m}$} ($E=-E_q+V_0=E_p$ or $q=\sqrt{(V_0-E)^2-m^2}$), the solution may be written as
\begin{equation}
\phi^{(E)}(z,t)={1\over\sqrt{2\pi}}\sqrt{m\over V_0-E}\left[\theta(-z)T\sigma_zu(-p,s)e^{-ipz}+\theta(z)\bigl\{v(-q,s)e^{iqz}+Rv(q,s)e^{-iqz}\bigr\}\right]e^{-iEt}, 
\end{equation}
with  
\begin{equation}
R=-{\sqrt{E+m\over V_0-E-m}-\sqrt{E-m\over V_0-E+m}\over\sqrt{E+m\over V_0-E-m}+\sqrt{E-m\over V_0-E+m}},\quad
T=-{2\over\sqrt{E+m\over V_0-E-m}+\sqrt{E-m\over V_0-E+m}}.
\end{equation}
The continuity of current and the conservation of probability are the same as in (\ref{eq:cccp2}).

\noindent
{\bf B3: for $\bm{m>E>-m}$}, no oscillating solution exists for the region $z<0$ and we have ($p=i\kappa=i\sqrt{m^2-E^2}$)
\begin{equation}
\phi^{(E)}(z,t)={1\over\sqrt{2\pi}}\sqrt{m\over V_0-E}\left[\theta(-z)T\sigma_zu(-i\kappa,s)e^{\kappa z}+\theta(z)\bigl\{v(-q,s)e^{iqz}+Rv(q,s)e^{-iqz}\bigr\}\right]e^{-iEt}, 
\end{equation}
for $E>0$ and
\begin{equation}
\phi^{(E)}(z,t)={1\over\sqrt{2\pi}}\sqrt{m\over V_0-E}\left[-i\theta(-z)Tv(i\kappa,s)e^{\kappa z}+\theta(z)\bigl\{v(-q,s)e^{iqz}+Rv(q,s)e^{-iqz}\bigr\}\right]e^{-iEt}, 
\end{equation}
for $E<0$, with  
\begin{equation}
R=-{\sqrt{m+E\over V_0-E-m}-i\sqrt{m-E\over V_0-E+m}\over\sqrt{m+E\over V_0-E-m}+i\sqrt{m-E\over V_0-E+m}},\quad
T=-{2\over\sqrt{m+E\over V_0-E-m}+i\sqrt{m-E\over V_0-E+m}}.
\end{equation}
In this case, all particles incident from the right are reflected
\begin{equation}
P_t=0,\quad P_r=|R|^2=1.
\end{equation}

\noindent
{\bf B4: for $\bm{-m>E}$}, $E=-E_p=-E_q+V_0=-|E|$, and we have (with positive $p$ and $q$)
\begin{equation}
\phi^{(E)}(z,t)={1\over\sqrt{2\pi}}\sqrt{m\over V_0-E}\left[\theta(-z)Tv(-p,s)e^{ipz}+\theta(z)\bigl\{v(-q,s)e^{iqz}+Rv(q,s)e^{-iqz}\bigr\}\right]e^{i|E|t}, 
\end{equation}
with
\begin{equation}
R={\sqrt{|E|+m\over|E|+V_0+m}-\sqrt{|E|-m\over|E|+V_0-m}\over\sqrt{|E|+m\over|E|+V_0+m}+\sqrt{|E|-m\over|E|+V_0-m}},\quad
T={2\over\sqrt{|E|+m\over|E|+V_0+m}+\sqrt{|E|-m\over|E|+V_0-m}}.
\end{equation}
In this case, the same current continuity relation and probability conservation as in (\ref{eq:cccp2}) follow.

\section{Orthonormality and completeness relation}
\label{sec:ProofComplete}
These eigenfunctions form a complete orthonormal set.
First, the orthogonality of the above scattering sates is due to the general argument for self-adjoint Hamiltonians, that is, the eigenfunctions belonging to different eigenvalues are orthogonal to each other.
Two $\psi$'s with different energies are orthogonal and normalized as
\begin{equation}
\int _{-\infty}^\infty dz\psi^{(E)}(z)^\dagger\psi^{(E')}(z)=\theta(EE')\delta(p-p'),
\label{eq:psidpsi}
\end{equation}
where $|E|=E_p,\,|E'|=E_{p'}$.
(The factor $\delta_{s,s'}$ representing the orthogonality in spin space is and will be suppressed but understood on the right-hand side and where it is necessary.)
Similarly, we should have 
\begin{equation}
\int _{-\infty}^\infty dz\phi^{(E)}(z)^\dagger\phi^{(E')}(z)=\theta\bigl[(E-V_0)(E'-V_0)\bigr]\delta(q-q'),
\label{eq:phidphi}
\end{equation}
where $|E-V_0|=E_q,\,|E'-V_0|=E_{q'}$ and
\begin{equation}
\int _{-\infty}^\infty dz\psi^{(E)}(z)^\dagger\phi^{(E')}(z)=0.
\label{eq:psidphi}
\end{equation}
Since the last orthogonality relation is not derived from the above argument for both $\psi^{(E)}$ and $\phi^{(E')}$ can happen to belong to the same energy $E=E'$, its validity has to be explicitly shown.
Incidentally, the fact that the above wave functions in the preceding section are properly normalized can be shown explicitly by calculating the left-hand sides of (\ref{eq:psidpsi}) and (\ref{eq:phidphi}) for each case, however, we may confirm it indirectly when they are shown to satisfy the completeness relation with the right factors.
The orthonormality conditions imply that the following form of completeness relation holds
\begin{equation}
\sum_{s,r=\pm}\int_0^\infty dp\,\psi_r^{(p)}(z)\psi_r^{(p)}{}^\dagger(z')+\sum_{s,r=\pm}\int_0^\infty dq\,\phi_r^{(q)}(z)\phi_r^{(q)}{}^\dagger(z')=\delta(z-z'),
\label{eq:complete}
\end{equation}
where the unit matrix in spinor space $\openone_{4\times4}$ has been suppressed on the right-hand side and we have introduced shorthand notations $\psi_\pm^{(p)}=\psi^{(\pm E_p)}$ and $\phi_\pm^{(q)}=\phi^{(\pm E_q+V_0)}$.
(The dependence of the eigenfunctions on the spin $s$ is suppressed as before even though the summation over spin is explicit here.)

\subsection{Proof of  the completeness relation: momentum integrations}
\label{subsec:proof}
As described in Introduction, though it is generally taken for granted that eigenfunctions of a Hamiltonian constitute a complete orthonormal set, which can be shown explicitly when it has only a finite number of discrete eigenstates, to prove it for a Hamiltonian endowed with a continuous spectrum can be another, nontrivial task.
In the present case, we need to show that the left-hand side of (\ref{eq:complete}) is diagonal in both the coordinate and spinor spaces, which would make the proof more involved than the non-relativistic cases \cite{TrottTrottSchnittler1989}. 
 
 Since the step potential takes different values depending on the sign of the coordinate $z$, we need to consider three cases, that is, $z, z'<0$, $0<z, z'$ and $z<0<z'$, separately to prove (\ref{eq:complete}).

 \noindent
 {\bf Case 1: $\bm{z, z'<0}$}
 
 \noindent
 When both $z$ and $z'$ are negative, i.e., on the left of the potential, the left-hand side of the completeness relation (\ref{eq:complete}) is explicitly written down as
 \begin{align}
&\int_0^\infty{dp\over2\pi}{m\over E_p}\bigl[u(p)e^{ipz}+R_pu(-p)e^{-ipz}\bigr]\bigl[u(p)e^{ipz'}+R_pu(-p)e^{-ipz'}\bigr]^\dagger\nonumber\\
 &
 +\int_0^\infty{dp\over2\pi}{m\over E_p}\bigl[v(p)e^{-ipz}+R_pv(-p)e^{ipz}\bigr]\bigl[v(p)e^{-ipz'}+R_pv(-p)e^{ipz'}\bigr]^\dagger
 +\int_0^\infty{dq\over2\pi}{m\over E_q}T_qu(-p)e^{-ipz}\bigl[T_qu(-p)e^{-ipz'}\bigr]^\dagger\nonumber\\
 &
 +\int_0^{\sqrt{(V_0-m)^2-m^2}}{dq\over2\pi}{m\over E_q}T_q\sigma_zu(-p)e^{-ipz}\bigl[T_q\sigma_zu(-p)e^{-ipz'}\bigr]^\dagger\nonumber\\
 &
 +\int_{\sqrt{(V_0-m)^2-m^2}}^{\sqrt{V_0^2-m^2}}{dq\over2\pi}{m\over E_q}T_q\sigma_zu(-i\kappa)e^{\kappa z}\bigl[T_q\sigma_zu(-i\kappa)e^{\kappa z'}\bigr]^\dagger
 +\int_{\sqrt{V_0^2-m^2}}^{\sqrt{(V_0+m)^2-m^2}}{dq\over2\pi}{m\over E_q}T_qv(i\kappa)e^{\kappa z}\bigl[T_qv(i\kappa)e^{\kappa z'}\bigr]^\dagger\nonumber\\
 &
+\int_{\sqrt{(V_0+m)^2-m^2}}^\infty{dq\over2\pi}{m\over E_q}T_qv(-p)e^{ipz}\bigl[T_qv(-p)e^{ipz'}\bigr]^\dagger,
 \end{align}
 where it is understood that the summation over spin $s$ has to be taken (though not explicitly written down) and the subscripts $p$ and $q$ are used to distinguish quantities associated with the left-incident ($p$) case and the right-incident ($q$) case (with the same energy).
 It is to be noticed that the reflection amplitude for the left-incident case $R_p$ can be expressed as
 \begin{equation}
 R_p={\sqrt{E_q+m\over E_p+m}-\sqrt{E_q-m\over E_p-m}\over\sqrt{E_q+m\over E_p+m}+\sqrt{E_q-m\over E_p-m}}
 \end{equation}
 for all $p\ge0$ by a proper analytical continuation from a large $p\ge\sqrt{(V_0+m)^2-m^2}$ (or  $E_p\ge V_0+m$).
Similarly the transmission amplitude for the right-incident case $T_q$ can be understood as a proper analytic continuation of 
\begin{equation}
T_q={2\over\sqrt{E_p+m\over E_q+m}+\sqrt{E_p-m\over E_q-m}}.
\end{equation}
After the change of variables from $q$ to $p$ (or $\kappa$ when $V_0-m\ge E_q\ge V_0+m$), the above expression is simplified as
\begin{align}
&\int_0^\infty{dp\over2\pi}{m\over E_p}\Bigl\{\bigl(u(p)u(p)^\dagger+|R_p|^2v(-p)v(-p)^\dagger\bigr)e^{ip(z-z')}+\bigl(v(p)v(p)^\dagger+|R_p|^2u(-p)u(-p)^\dagger\bigr)e^{-ip(z-z')}\nonumber\\
&\qquad\qquad\quad
+\bigl(R_p^*u(p)u(-p)^\dagger+R_pv(-p)v(p)^\dagger\bigr)e^{ip(z+z')}+\bigl(R_pu(-p)u(p)^\dagger+R_p^*v(p)v(-p)^\dagger\bigr)e^{-ip(z+z')}\Bigr\}\nonumber\\
&+\int_0^\infty{dp\over2\pi}{m\over E_p}{p\over q}|T_q|^2u(-p)u(-p)^\dagger e^{-ip(z-z')}\nonumber\\
&+\int_0^m{d\kappa\over2\pi}{m\over E_{(\kappa)}}\Bigl({\kappa\over q}|T_q|^2u(-i\kappa)u(-i\kappa)^\dagger+{\kappa\over q}|T_q|^2v(i\kappa)v(i\kappa)^\dagger\Bigr)e^{\kappa(z+z')}\nonumber\\
&+\int_0^\infty{dp\over2\pi}{m\over E_p}{p\over q}|T_q|^2v(-p)v(-p)^\dagger e^{ip(z-z')},
\end{align}
where $E_{(\kappa)}=E_{\pm i\kappa}=\sqrt{m^2-\kappa^2}$ and the fact that $\sigma_z$ is commutable with both spinors $u$ and $v$ has been used.
We have to pay due attention to the fact that, even though not explicitly written, the values of $q$ (and therefore those of $E_q$) are different when it appears in association with the spinor $u$ or $v$, which, of course, applies also to $R_p$ and $T_q$ in the above.
(This is the reason why apparently the same quantities have not been put together in the last but one line.)
Observe that there are terms of  functions of  the difference $z-z'$ and the sum $z+z'$ and they are apparently independent of each other.

First, all terms that are functions of $z-z'$ are collected to yield
\begin{align}
\int_0^\infty{dp\over2\pi}{m\over E_p}\Bigl\{&\bigl(u(p)u(p)^\dagger+|R_p|^2v(-p)v(-p)^\dagger+{p\over q}|T_q|^2v(-p)v(-p)^\dagger\bigr)e^{ip(z-z')}\nonumber\\
&+\bigl(v(p)v(p)^\dagger+|R_p|^2u(-p)u(-p)^\dagger+{p\over q}|T_q|^2u(-p)u(-p)^\dagger\bigr)e^{-ip(z-z')}\Bigr\}.
\end{align}
We note that the transmission probability for the right-incident case ${p\over q}|T_q|^2$ is the same as that for the left-incident case ${q\over p}|T_p|^2$, which can be explicitly shown by direct calculation, stating that the reciprocal relation in quantum mechanics also holds in this case.
Finally, the conservation of probability $|R_p|^2+{q\over p}|T_p|^2=1$ greatly simplifies the above to reach
\begin{align}
\int_{-\infty}^\infty{dp\over2\pi}{m\over E_p}\bigl(u(p)u(p)^\dagger+v(-p)v(-p)^\dagger\bigr)e^{ip(z-z')}
&=\int_{-\infty}^\infty{dp\over2\pi}{m\over E_p}\Bigl({E_p\gamma^0-p\gamma^3+m\over2m}\gamma^0+{E_p\gamma^0+p\gamma^3-m\over2m}\gamma^0\Bigr)e^{ip(z-z')}\nonumber\\
&=\delta(z-z'),
\end{align}
where at the first equality the spin sum has explicitly been taken and the unit matrix $\openone_{4\times4}$ is suppressed on the right-most hand.

Second, the remaining terms that are functions of $z+z'$ are shown to cancel to each other.
We understand that the contributions coming from the $p$-integral
\begin{equation}
\int_0^\infty{dp\over2\pi}{m\over E_p}\Bigl\{\bigl(R_p^*u(p)u(-p)^\dagger+R_pv(-p)v(p)^\dagger\bigr)e^{ip(z+z')}+\bigl(R_pu(-p)u(p)^\dagger+R_p^*v(p)v(-p)^\dagger\bigr)e^{-ip(z+z')}\Bigr\}
\label{eq:RuuRvv}
\end{equation}
can be evaluated on the complex $p$-plane, because the combination $z+z'$ in the exponents is negative definite, the reflection amplitude decays at least like $1/|p|$ for large $|p|$, assuring the convergence of the integrands at $|p|\to\infty$  and the integrands have no singularities other than the several cuts on the real and imaginary $p$-axises.
 The integrand proportional to $e^{ip(z+z')}$ can be evaluated on the negative imaginary axis $p=-i\kappa$, while the other one proportional to $e^{-ip(z+z')}$ is to be evaluated on the positive imaginary axis $p=i\kappa$, where $\kappa\ge0$ in both cases.
On the negative imaginary $p$-axis, we set $p=-i\kappa$ and
\begin{equation}
u(p)u(-p)^\dagger e^{ip(z+z')}=u(-i\kappa)u(-i\kappa)^\dagger e^{\kappa(z+z')},\quad
v(-p)v(p)^\dagger e^{ip(z+z')}=v(i\kappa)v(i\kappa)^\dagger e^{\kappa(z+z')},
\end{equation}
while on the positive imaginary $p$-axis $p=i\kappa$, we have
\begin{equation}
u(-p)u(p)^\dagger e^{-ip(z+z')}=u(-i\kappa)u(-i\kappa)^\dagger e^{\kappa(z+z')},\quad
v(p)v(-p)^\dagger e^{-ip(z+z')}=v(i\kappa)v(i\kappa)^\dagger e^{\kappa(z+z')}.
\end{equation}

For small $\kappa\le m$, $E_p$ is a real number smaller than $m$, $0\le E_p\le m$, and the reflection amplitude associated with spinor $u$ is analytically continued as
\begin{align}
R_p^*&\to{\sqrt{E_p-V_0+m\over E_p+m}-\sqrt{E_p-V_0-m\over E_p-m}\over\sqrt{E_p-V_0+m\over E_p+m}+\sqrt{E_p-V_0-m\over E_p-m}}\Biggr\vert_{p=-i\kappa}={-i\sqrt{V_0-m-E\over E+m}-\sqrt{V_0+m-E\over m-E}\over-i\sqrt{V_0-m-E\over E+m}+\sqrt{V_0+m-E\over m-E}}\Biggr\vert_{E=E_{-i\kappa}}\nonumber\\
R_p&\to{\sqrt{E_p-V_0+m\over E_p+m}-\sqrt{E_p-V_0-m\over E_p-m}\over\sqrt{E_p-V_0+m\over E_p+m}+\sqrt{E_p-V_0-m\over E_p-m}}\Biggr\vert_{p=i\kappa}={i\sqrt{V_0-m-E\over E+m}-\sqrt{V_0+m-E\over m-E}\over i\sqrt{V_0-m-E\over E+m}+\sqrt{V_0+m-E\over m-E}}\Biggr\vert_{E=E_{i\kappa}},
\end{align}
because the purely imaginary quantity $\sqrt{E_p-V_0+m}$ has the same phase as that of $p$ on the imaginary axis, for
\begin{equation}
\sqrt{E_p-V_0+m}={\sqrt{E_p^2-(V_0-m)^2}\over\sqrt{E_p+V_0-m}}={\sqrt{p-p_0}\sqrt{p+p_0}\over\sqrt{E_p+V_0-m}},
\end{equation}
where $p_0=\sqrt{(V_0-m)^2-m^2}$ is a real number and located on the real axis.
Similarly, for the reflection amplitude associated with spinor $v$, we obtain
\begin{align}
R_p&\to{\sqrt{E_p+V_0+m\over E_p+m}-\sqrt{E_p+V_0-m\over E_p-m}\over\sqrt{E_p+V_0+m\over E_p+m}+\sqrt{E_p+V_0-m\over E_p-m}}\Biggr\vert_{p=-i\kappa}={\sqrt{E+V_0+m\over E+m}-i\sqrt{E+V_0-m\over m-E}\over\sqrt{E+V_0+m\over E+m}+i\sqrt{E+V_0-m\over m-E}}\Biggr\vert_{E=E_{-i\kappa}}\nonumber\\
R_p^*&\to{\sqrt{E_p+V_0+m\over E_p+m}-\sqrt{E_p+V_0-m\over E_p-m}\over\sqrt{E_p+V_0+m\over E_p+m}+\sqrt{E_p+V_0-m\over E_p-m}}\Biggr\vert_{p=i\kappa}={\sqrt{E+V_0+m\over E+m}+i\sqrt{E+V_0-m\over m-E}\over\sqrt{E+V_0+m\over E+m}-i\sqrt{E+V_0-m\over m-E}}\Biggr\vert_{E=E_{i\kappa}}.
\end{align}
All contributions coming from small $\kappa$ on the imaginary axis are put together to yield
\begin{equation}
\int_0^m{d\kappa\over2\pi}{m\over E_{(\kappa)}}\Bigl\{\Bigl({\kappa\over q}|T_q|^2+i(R_p-R_p^*)\Bigr)u(-i\kappa)u(-i\kappa)^\dagger+\Bigl({\kappa\over q}|T_q|^2-i(R_p-R_p^*)\Bigr)v(i\kappa)v(i\kappa)^\dagger\Bigr\}e^{\kappa(z+z')}.
\end{equation}
The coefficient of spinor $u$ vanishes, because
\begin{align}
&{\kappa\over q}\left|-{2\over\sqrt{m+E\over V_0-E-m}+i\sqrt{m-E\over V_0-E+m}}\right|^2_{E=E_{(\kappa)}}-\left.2\Im{i\sqrt{V_0-E-m\over m+E}-\sqrt{V_0-E+m\over m-E}\over i\sqrt{V_0-E-m\over m+E}+\sqrt{V_0-E+m\over m-E}}\right\vert_{E=E_{(\kappa)}}\nonumber\\
&={4\kappa\over q}\left({1\over{m+E\over V_0-E-m}+{m-E\over V_0-E+m}}-{1\over{\kappa^2\over q^2}({V_0-E-m\over m+E}+{V_0-E+m\over m-E})}\right)_{E=E_{(\kappa)}}=0,
\end{align}
where we have used the relations $\kappa^2=m^2-E^2$ and $q^2=(V_0-E)^2-m^2$ with $E=E_{(\kappa)}$.
Similarly, the coefficient of spinor $v$, since in this case $q^2=(V_0+E)^2-m^2$ with $E=E_{(\kappa)}$, also vanishes
\begin{align}
&{\kappa\over q}\left|-{2\over\sqrt{m-E\over V_0+E-m}+i\sqrt{m+E\over V_0+E+m}}\right|^2_{E=E_{(\kappa)}}+\left.2\Im{\sqrt{V_0+E+m\over m+E}-i\sqrt{V_0+E-m\over m-E}\over\sqrt{V_0+E+m\over m+E}+i\sqrt{V_0+E-m\over m-E}}\right\vert_{E=E_{(\kappa)}}\nonumber\\
&={4\kappa\over q}\left({1\over{m-E\over V_0+E-m}+{m+E\over V_0+E+m}}-{1\over{\kappa^2\over q^2}({V_0+E+m\over m+E}+{V_0+E-m\over m-E})}\right)_{E=E_{(\kappa)}}=0.
\end{align}

Thus what is remained is the contributions arising from large $\kappa\ge m$, where $E_p$ becomes purely imaginary.
First we note that the function $E_p$ gains the same phase as that of $p$ on the imaginary axis, i.e., $E_{\pm i\kappa}=\pm i\sqrt{\kappa^2-m^2}\equiv\pm i\Ek$ for $\kappa\ge m$.
From the direct calculations, we understand that the reflection amplitude for spinor $u$ is replaced by the following complex numbers on the imaginary axis
\begin{equation}
R_p\to{\sqrt{i\Ek-V_0+m\over i\Ek+m}-\sqrt{i\Ek-V_0-m\over i\Ek-m}\over\sqrt{i\Ek-V_0+m\over i\Ek+m}+\sqrt{i\Ek-V_0-m\over i\Ek-m}}= R_u(i\Ek),\quad
R_p^*\to{\sqrt{-i\Ek-V_0+m\over-i\Ek+m}-\sqrt{-i\Ek-V_0-m\over-i\Ek-m}\over\sqrt{-i\Ek-V_0+m\over-i\Ek+m}+\sqrt{-i\Ek-V_0-m\over-i\Ek-m}} =R_u(-i\Ek),
\end{equation}
while for spinor $v$,
\begin{equation}
R_p\to{\sqrt{-i\Ek+V_0+m\over-i\Ek+m}-\sqrt{-i\Ek+V_0-m\over-i\Ek-m}\over\sqrt{-i\Ek+V_0+m\over-i\Ek+m}+\sqrt{-i\Ek+V_0-m\over-i\Ek-m}}= R_v(-i\Ek),\quad
R_p^*\to{\sqrt{i\Ek+V_0+m\over i\Ek+m}-\sqrt{i\Ek+V_0-m\over i\Ek-m}\over\sqrt{i\Ek+V_0+m\over i\Ek+m}+\sqrt{i\Ek+V_0-m\over i\Ek-m}}=R_v(i\Ek).
\end{equation}
Observe that the following equalities hold
\begin{equation}
R_v(i\Ek)=-R_u(-i\Ek),\quad R_v(-i\Ek)=-R_u(i\Ek).
\label{eq:RuRv}
\end{equation}
The spinor parts for $u$ and $v$ are explicitly worked out, after summing over spin degrees freedom, to be
\begin{align}
u(-i\kappa)u(-i\kappa)^\dagger&={\sqrt{-i\Ek+m}\sqrt{i\Ek+m}\over2m}
\begin{pmatrix}1\\{-i\sigma_z\kappa\over-i\Ek+m}\end{pmatrix}
\left(1,\,{i\sigma_z\kappa\over i\Ek+m}\right)={1\over2m}\begin{pmatrix}\kappa&i\sigma_z(-i\Ek+m)\\-i\sigma_z(i\Ek+m)&\kappa\end{pmatrix},\nonumber\\
v(i\kappa)v(i\kappa)^\dagger&={\sqrt{i\Ek+m}\sqrt{-i\Ek+m}\over2m}
\begin{pmatrix}{i\sigma_z\kappa\over i\Ek+m}\\1\end{pmatrix}
\left({-i\sigma_z\kappa\over-i\Ek+m},\,1\right)={1\over2m}\begin{pmatrix}\kappa&i\sigma_z(-i\Ek+m)\\-i\sigma_z(i\Ek+m)&\kappa\end{pmatrix},
\label{eq:uud=vvd}
\end{align}
where the relation $-\Ek^2=m^2-\kappa^2$ has been used, showing explicitly that they are actually the same.
The remaining integrations over $\kappa$ are now written as
\begin{align}
&-i\int_m^\infty{d\kappa\over2\pi}{m\over-i\Ek}
\left[R_u(-i\Ek)u(-i\kappa)u(-i\kappa)^\dagger+R_v(-i\Ek)v(i\kappa)v(i\kappa)^\dagger\right]e^{\kappa(z+z')}\nonumber\\
&\quad
+i\int_m^\infty{d\kappa\over2\pi}{m\over i\Ek}
\left[R_u(i\Ek)u(-i\kappa)u(-i\kappa)^\dagger+R_v(i\Ek)v(i\kappa)v(i\kappa)^\dagger\right]e^{\kappa(z+z')},
\end{align}
which vanishes owing to the above relations (\ref{eq:RuRv}) and (\ref{eq:uud=vvd}).
This completes the proof of the completeness relation (\ref{eq:complete}) for $z,z'<0$.

\noindent
{\bf Case 2: $\bm{z,z'>0}$}

\noindent
As in the previous case, the left-hand side of (\ref{eq:complete}) is explicitly written down
\begin{align}
&\int_{\sqrt{(V_0+m)^2-m^2}}^\infty{dp\over2\pi}{m\over E_p}T_pu(q)e^{iqz}\bigl[T_pu(q)e^{ipz'}\bigr]^\dagger\nonumber\\
&
+\int_{\sqrt{V_0^2-m^2}}^{\sqrt{(V_0+m)^2-m^2}}{dp\over2\pi}{m\over E_p}T_pu(i\kappa')e^{-\kappa' z}\bigl[T_pu(i\kappa')e^{-\kappa' z'}\bigr]^\dagger
+\int_{\sqrt{(V_0-m)^2-m^2}}^{\sqrt{V_0^2-m^2}}{m\over E_p}T_pi\sigma_z v(-i\kappa')e^{-\kappa' z}\bigl[T_pi\sigma_z v(-i\kappa')e^{-\kappa' z'}\bigr]^\dagger\nonumber\\
&
+\int_0^{\sqrt{(V_0-m)^2-m^2}}{dp\over2\pi}{m\over E_p}T_p\sigma_z v(q)e^{-iqz}\bigl[T_p\sigma_z v(q)e^{-iqz'}\bigr]^\dagger\nonumber\\
&
+\int_0^\infty{dp\over2\pi}{m\over E_p}T_pv(q)e^{-iqz}\bigl[T_pv(q)e^{-iqz'}\bigr]^\dagger
+\int_0^\infty{dq\over2\pi}{m\over E_q}\bigl[u(-q)e^{-iqz}+R_qu(q)e^{iqz}\bigr]\bigl[u(-q)e^{-iqz'}+R_qu(q)e^{iqz'}\bigr]^\dagger\nonumber\\
&
+\int_0^\infty{dq\over2\pi}{m\over E_q}\bigl[v(-q)e^{iqz}+R_qv(q)e^{-iqz}\bigr]\bigl[v(-q)e^{iqz'}+R_qv(q)e^{-iqz'}\bigr]^\dagger,
\end{align}
where $\kappa'\ge0$ is defined, when $E_q\le m$, as $E_q^2=m^2-\kappa'^2$.
Change of variables $p\to q$ (or $\kappa'$) reduces this to 
\begin{align}
&\int_0^\infty{dq\over2\pi}{m\over E_q}\Biggl\{\Bigl(\bigl({q\over p}|T_p|^2+|R_q|^2\bigr)u(q)u(q)^\dagger+v(-q)v(-q)^\dagger\Bigr)e^{iq(z-z')}\nonumber\\
&\qquad\qquad\quad
+\Bigl(\bigl({q\over p}|T_p|^2+|R_q|^2\bigr)v(q)v(q)^\dagger+u(-q)u(-q)^\dagger\Bigr)e^{-iq(z-z')}\nonumber\\
&\qquad\qquad\quad
+\bigl(R_q^*u(-q)u(q)^\dagger+R_qv(q)v(-q)^\dagger\bigr)e^{-iq(z+z')}+\bigl(R_qu(q)u(-q)^\dagger+R_q^*v(-q)v(q)^\dagger\bigr)e^{iq(z+z')}\Biggr\}\nonumber\\
&
+\int_0^m{d\kappa'\over2\pi}{m\over E_p}\Bigl({\kappa'\over p}|T_p|^2u(i\kappa')u(i\kappa')^\dagger+{\kappa'\over p}|T_p|^2v(-i\kappa')v(-i\kappa')^\dagger\Bigr)e^{-\kappa'(z+z')}.
\label{eq:++}
\end{align}
The reciprocal relation ${q\over p}|T_p|^2={p\over q}|T_q|^2$ also holds true  in this case, which can be proven by direct calculation, and the probability conservation $|R_q|^2+{p\over q}|T_q|^2=1$ brings about the delta function $\delta(z-z')$ from those terms that depend on the difference $z-z'$ in the above (\ref{eq:++}).
The remaining integrations over $q$ are evaluated on the imaginary $q$-axis, where spinors become $u(i\kappa')u(i\kappa')^\dagger$ and $v(-i\kappa')v(-i\kappa')^\dagger$ and the reflection amplitude $R_q$ for spinor $u$ is represented as
\begin{align}
R_q={\sqrt{E_q+V_0+m\over E_q+m}-\sqrt{E_q+V_0-m\over E_q-m}\over\sqrt{E_q+V_0+m\over E_q+m}+\sqrt{E_q+V_0-m\over E_q-m}}&\to
{\sqrt{E_q+V_0+m\over E_q+m}+i\sqrt{E_q+V_0-m\over m-E_q}\over\sqrt{E_q+V_0+m\over E_q+m}-i\sqrt{E_q+V_0-m\over m-E_q}}\quad{\rm for\ }\kappa'\le m,\,E_q=\sqrt{m^2-\kappa'^2}\nonumber\\
&\to{\sqrt{i\Ekp+V_0+m\over i\Ekp+m}-\sqrt{i\Ekp+V_0-m\over i\Ekp-m}\over\sqrt{i\Ekp+V_0+m\over i\Ekp+m}+\sqrt{i\Ekp+V_0-m\over i\Ekp-m}}=R_u(i\Ekp)\quad{\rm for\ }\kappa'\ge m,\,\Ekp=\sqrt{\kappa'^2-m^2},
\end{align}
and for $v$,
\begin{align}
R_q={\sqrt{E_q-V_0+m\over E_q+m}-\sqrt{E_q-V_0-m\over E_q-m}\over\sqrt{E_q-V_0+m\over E_q+m}+\sqrt{E_q-V_0-m\over E_q-m}}&\to
{-i\sqrt{V_0-E_q-m\over E_q+m}-\sqrt{V_0-E_q+m\over m-E_q}\over-i\sqrt{V_0-E_q-m\over E_q+m}+\sqrt{V_0-E_q+m\over m-E_q}}\quad{\rm for\ }\kappa'\le m,\,E_q=\sqrt{m^2-\kappa'^2}\nonumber\\
&\to{\sqrt{-i\Ekp-V_0+m\over-i\Ekp+m}-\sqrt{-i\Ekp-V_0-m\over -i\Ekp-m}\over\sqrt{-i\Ekp-V_0+m\over -i\Ekp+m}+\sqrt{-i\Ekp-V_0-m\over-i\Ekp-m}}=R_v(-i\Ekp)\quad{\rm for\ }\kappa'\ge m,\,\Ekp=\sqrt{\kappa'^2-m^2}
\end{align}
and $R_q^*$'s as the complex conjugates of the corresponding ones.
Thus the contributions of terms that are functions of $z+z'$ are
\begin{align}
&\int_0^m{d\kappa'\over2\pi}{m\over E_q}\Bigl\{\Bigl({\kappa'\over p}|T_p|^2+i(R_q-R_q^*)\Bigr)u(i\kappa')u(i\kappa')^\dagger+\Bigl({\kappa'\over p}|T_p|^2-i(R_q-R_q^*)\Bigr)v(-i\kappa')v(-i\kappa')^\dagger\Bigr\}e^{-\kappa'(z+z')}\nonumber\\
&
-i\int_m^\infty{d\kappa'\over2\pi}{m\over-i\Ekp}
\left[R_u(-i\Ekp)u(i\kappa')u(i\kappa')^\dagger+R_v(-i\Ekp)v(-i\kappa')v(-i\kappa')^\dagger\right]e^{-\kappa'(z+z')}\nonumber\\
&
+i\int_m^\infty{d\kappa'\over2\pi}{m\over i\Ekp}
\left[R_u(i\Ekp)u(i\kappa')u(i\kappa')^\dagger+R_v(i\Ekp)v(-i\kappa')v(-i\kappa')^\dagger\right]e^{-\kappa'(z+z')},
\label{eq:RuuRvv2}
\end{align}
which can be shown to vanish on the basis of the similar arguments as before.
This completes the proof for $z,z'>0$. 

\noindent
{\bf Case 3: $\bm{z<0<z'}$}

\noindent
If the left-hand side of (\ref{eq:complete}) is explicitly written down in this case, there are eight (actually ten) different terms existing, corresponding to eight different cases {\bf A1}$\sim${\bf A4} and {\bf B1}$\sim${\bf B4}.
It can be shown, however, that these terms are categorized into four groups, according to their energy eigenvalues, i) $E\ge m$, ii) $m>E>0$, iii) $0>E>-m$ and iv) $-m\ge E$.

\noindent
i) $E\ge m$: Consider first the contributions arising from those wave functions that are belonging to eigenvalues (energies) greater than $V_0+m$.
They are expressed as
\begin{align}
&\int_{E=E_p\ge V_0+m}^\infty{dp\over2\pi}{m\over E_p}\bigl[u(p)e^{ipz}+R_pu(-p)e^{-ipz}\bigr]\bigl[T_pu(q)e^{iqz'}\bigr]^\dagger
+\int_0^\infty{dq\over2\pi}{m\over E_q}T_qu(-p)e^{-ipz}\bigl[u(-q)^{-iqz'}+R_qu(q)e^{iqz'}\bigr]^\dagger \nonumber\\
&=\int_{\sqrt{(V_0+m)^2-m^2}}^\infty{dp\over2\pi}{m\over E_p}\Bigl\{T_p^*u(p)u(q)^\dagger e^{ipz-iqz'}+R_pT_p^*u(-p)u(q)^\dagger e^{-ipz-iqz'}\nonumber\\
&\qquad\qquad\qquad\qquad\quad
+{p\over q}\Bigl(T_qu(-p)u(-q)^\dagger e^{-ipz+iqz'}+T_qR_q^*u(-p)u(q)^\dagger e^{-ipz-iqz'}\Bigr)\Bigr\}.
\label{eq:Ep>V0+m}
\end{align}
Observe that in this energy range the reflection and transmission amplitudes are all real-valued and, moreover, the following relations hold
\begin{equation}
{p\over q}T_q={2\over\bigl({q\over p}\bigr)\Bigl(\sqrt{E_p+m\over E_q+m}+\sqrt{E_p-m\over E_q-m}\Bigr)}={2\over\sqrt{E_q-m\over E_p-m}+\sqrt{E_q+m\over E_p+m}}=T_p
\end{equation}
and
\begin{equation}
{p\over q}R_q^*T_q={2\bigl({q\over p}\bigr)\Bigl(\sqrt{E_p+m\over E_q+m}-\sqrt{E_q-m\over E_p-m}\Bigr)\over{
\bigl({q\over p}\bigr)^2\Bigl(\sqrt{E_p+m\over E_q+m}+\sqrt{E_q-m\over E_p-m}\Bigr)^2}}
={2\Bigl(\sqrt{E_q-m\over E_p-m}-\sqrt{E_q+m\over E_p+m}\Bigr)\over\Bigl(\sqrt{E_q-m\over E_p-m}+\sqrt{E_q+m\over E_p+m}\Bigr)^2}=-R_pT_p^*.
\end{equation}
Thus the above expression (\ref{eq:Ep>V0+m}) is simplified to 
\begin{equation}
\int_{\sqrt{(V_0+m)^2-m^2}}^\infty{dp\over2\pi}{m\over E_p}\bigl(T_p^*u(p)u(q)^\dagger e^{ipz-iqz'}+T_pu(-p)u(-q)^\dagger e^{-ipz+iqz'}\bigr).
\label{eq:Elarge}
\end{equation}

As a matter of fact, it can be confirmed that the contributions coming from the wave functions belonging to energy eigenvalues greater than or equal to $m$, $E\ge m$, are consistently written in this form with the lower limit replaced with $p=0$.
When $\sqrt{(V_0+m)^2-m^2}>p\ge\sqrt{V_0^2-m^2}$, in order to make the integrand convergent at $|p|\to\infty$, the first term in (\ref{eq:Elarge}) is defined by 
\begin{equation}
T_p^*u(p)u(q)^\dagger e^{ipz-iqz'}\Bigr\vert_{q\to-i\kappa'}=T_p^*u(p)u(i\kappa')^\dagger e^{ipz-\kappa'z'},
\end{equation}
while the second term is defined as 
\begin{equation}
T_pu(-p)u(-q)^\dagger e^{-ipz+iqz'}\Bigr\vert_{q\to i\kappa'}=T_pu(-p)u(i\kappa')^\dagger e^{-ipz-\kappa'z'}.
\end{equation}
Notice that here on the left-hand side, the variable $q$ is to be replaced with that specified at vertical bar $-i\kappa'$ or $i\kappa'$ in the quantities which are considered as functions of $q$, while on the right-hand side, conjugate operations are to be taken for functions of specified variables.
For example, $u(q)|_{q\to i\kappa'}=u(i\kappa')$ and $u(q)^\dagger|_{q\to i\kappa'}=u(-i\kappa')^\dagger$.
Since, in the second term of (\ref{eq:Elarge}), we analytically continue $q\to i\kappa'$, the term  $\sqrt{E_q-m}=\sqrt{E_p-V_0-m}$ is replaced with $i\sqrt{V_0-E_p+m}$ in the amplitudes $T_p$, which results in the relation $T_p=R_pT_p^*$.
(This relation also holds when $V_0>E\ge V_0-m$.)
The sum of these two terms is just the contribution from the energy range $V_0\le E<V_0+m$.

In the energy range $V_0-m\le E<V_0$, we note that $\kappa'=\sqrt{(E_p-V_0+m)(m-E_p+V_0)}$ and 
\begin{align}
u(i\kappa')^\dagger&=\sqrt{E_p-V_0+m\over2m}\Bigl(1,\,{-i\sigma_z\kappa'\over E_p-V_0+m}\Bigr)={1\over\sqrt{2m}}\Bigl(\sqrt{E_p-V_0+m},\,-i\sigma_z\sqrt{m-E_p+V_0}\Bigr)\nonumber\\
&=-i{1\over\sqrt{2m}}\Bigl(i\sigma_z\sqrt{E_p-V_0+m},\,\sqrt{m-E_p+V_0}\Bigr)\sigma_z=\sqrt{V_0-E_p+m\over2m}\Bigl({i\sigma_z\kappa'\over V_0-E_p+m},\,1\Bigr)\sigma_z=-iv(-i\kappa')^\dagger\sigma_z.
\end{align}
Therefore, the integrand in the parentheses in (\ref{eq:Elarge}) becomes nothing but
\begin{equation}
(-i)T_p^*u(p)v(-i\kappa')^\dagger\sigma_ze^{ipz-\kappa'z'}+(-i)R_pT_p^*u(-p)v(-i\kappa')^\dagger\sigma_ze^{-ipz-\kappa'z'},
\end{equation}
which is the contribution from this energy range $V_0-m\le E<V_0$.

Finally in the energy range $m\le E<V_0-m$, the variable $q$ is a positive number and appears in association with the spinor $v$ as $v(q)e^{-iqz'}$ or $v(-q)e^{iqz'}$, which implies that the term $\sqrt{E-V_0\pm m}$ has to be replaced with $i\sqrt{V_0-E\mp m}$ in the amplitude $T_p$ and with $-i\sqrt{V_0-E\mp m}$ in $T_p^*$ in (\ref{eq:Elarge}).
 Note that in either case, $q$ is to be replaced with $-q$.
It is straightforward to confirm that the following relations hold, for the left-incident contribution, 
\begin{align}
&\int_0^{\sqrt{(V_0-m)^2-m^2}}{dp\over2\pi}{m\over E_p}\bigl[u(p)e^{ipz}+R_pu(-p)e^{-ipz}\bigr]\bigl[T_p\sigma_zv(q)e^{-iqz'}\bigr]^\dagger\nonumber\\
&=\int_0^{\sqrt{(V_0-m^2)-m^2}}{dp\over2\pi}{m\over E_p}\bigl[u(p)e^{ipz}+R_pu(-p)e^{-ipz}\bigr]\bigl[T_pu(q)e^{iqz'}\bigr]^\dagger\Bigr\vert_{\sqrt{E_p-V_0\pm m}\to-i\sqrt{V_0-E_p\mp m}},
\end{align}
and for the right-incident contribution,
\begin{align}
&\int_0^{\sqrt{(V_0-m)^2-m^2}}{dp\over2\pi}{m\over E_p}{p\over q}T_q\sigma_zu(-p)e^{-ipz}\bigl[v(-q)e^{iqz'}+R_qv(q)e^{-iqz'}\bigr]^\dagger\nonumber\\
&=\int_0^{\sqrt{(V_0-m)^2-m^2}}{dp\over2\pi}{m\over E_p}T_pu(-p)e^{-ipz}\bigl[u(-q)e^{-iqz'}-R_pu(q)e^{iqz'}\bigr]^\dagger\Bigr\vert_{\sqrt{E_p-V_0\pm m}\to i\sqrt{V_0-E_p\mp m}}.
\end{align}
The second terms on the right-hand sides of these relations (i.e., those proportional to the product of reflection and transmission amplitudes) are canceled to each other and finally we arrive at the conclusion that the contributions coming from the energy range $E\ge m$ are consistently written down as
\begin{equation}
\int_0^\infty{dp\over2\pi}{m\over E_p}\bigl(T_p^*u(p)u(q)^\dagger e^{ipz-iqz'}+T_pu(-p)u(-q)^\dagger e^{-ipz+iqz'}\bigr).
\label{eq:Egtm}
\end{equation}
\noindent
ii) $m>E>0$: 
A straightforward calculation yields
\begin{align}
&\int_{\sqrt{(V_0-m)^2-m^2}}^{\sqrt{V_0^2-m^2}}{dq\over2\pi}{m\over E_q}T_q\sigma_zu(-i\kappa)e^{\kappa z}\bigl[v(-q)e^{iqz'}+R_qv(q)e^{-iqz'}\bigr]^\dagger\nonumber\\
&=\int_0^m{d\kappa\over2\pi}{\kappa\over q}{m\over E_p}\Bigl(T_q\sigma_zu(-i\kappa)v(-q)^\dagger e^{\kappa z-iqz'}-T_q^*\sigma_zu(-i\kappa)v(q)^\dagger e^{\kappa z+iqz'}\Bigr)\nonumber\\
&=-\int_0^{im}{d(i\kappa)\over2\pi}{m\over E_p}T_pu(-p)u(-q)^\dagger e^{-ipz+iqz'}\Bigr\vert_{p\to i\kappa}
-\int_0^{-im}{d(-i\kappa)\over2\pi}{m\over E_p}T_p^*u(p)u(q)^\dagger e^{ipz-iqz'}\Bigr\vert_{p\to-i\kappa},
\end{align}
where $T_qR_q^*=-T_q^*$ has been used in the first equality.
Actually in the first term of the last line, $T_p$ is evaluated by replacing $\sqrt{E_p-V_0\pm m}$ by $i\sqrt{V_0-E_p\mp m}$ in the original one in (\ref{eq:originalRT+}) and it is confirmed that it coincides with $-{\kappa\over q}T_q$.
Under these replacements together with $p\to i\kappa$, we can show that 
\begin{align}
u(-q)^\dagger&=\sqrt{E_p-V_0+m\over2m}\Bigl(1,\,-\sigma_z\sqrt{E_p-V_0-m\over E_p-V_0+m}\Bigr)\nonumber\\
&\to i\sqrt{V_0-E_p-m\over2m}\Bigl(1,-\sigma_z\sqrt{E_p-V_0-m\over E_p-V_0+m}\Bigr)
=-i\sqrt{V_0-E_p+m\over2m}\Bigl(-\sigma_z\sqrt{V_0-E_p-m\over V_0-E_p+m},\,1\Bigr)\sigma_z=-iv(-q)^\dagger\sigma_z.
\end{align}
(Remember that under these replacements, $q$ is to be replaced with $-q$ and we get the right exponential factors.)
A similar argument is applied to the second term.
The contribution is now summarized as
\begin{equation}
\int_{-im}^0{dp\over2\pi}{m\over E_p}T_p^*u(p)u(q)^\dagger e^{ipz-iqz'}+\int_{im}^0{dp\over2\pi}{m\over E_p}T_pu(-p)u(-q)^\dagger e^{-ipz+iqz'},
\end{equation}
where the integration contours are taken along the imaginary $p$-axis.

\noindent
iii) \& iv) $0\ge E$:
We first consider the region iv) where $E\le-m$.
The contribution coming from this energy range is
\begin{align}
&\int_0^\infty{dp\over2\pi}{m\over E_p}\bigl[v(p)e^{-ipz}+R_pv(-p)e^{ipz}\bigr]\bigl[T_pv(q)e^{-iqz'}\bigr]^\dagger
+\int_{\sqrt{(V_0+m)^2-m^2}}^\infty{dq\over2\pi}{m\over E_q}T_qv(-p)e^{ipz}\bigl[v(-q)e^{iqz'}+R_qv(q)e^{-iqz'}\bigr]^\dagger\nonumber\\
&=\int_0^\infty{dp\over2\pi}{m\over E_p}\Bigl(T_p^*v(p)v(q)^\dagger e^{-ipz+iqz'}+T_pv(-p)v(-q)^\dagger e^{ipz-iqz'}\Bigr).
\end{align}
 Since in the contribution arising from the region iii) $0>E>-m$
 \begin{equation}
 \int_{\sqrt{V_0^2-m^2}}^{\sqrt{(V_0+m)^2-m^2}}{dq\over2\pi}{m\over E_q}T_q(-i)v(i\kappa)e^{\kappa z}\bigl[v(-q)e^{iqz'}+R_qv(q)e^{-iqz'}\bigr]^\dagger
 \end{equation}
 the relation $T_qR_q^*=-T_q^*$ holds, we can rewrite this as
 \begin{equation}
 -\int_m^0{d\kappa\over2\pi}{m\over E_p}{\kappa\over q}\Bigl(-iT_qv(i\kappa)v(-q)^\dagger e^{\kappa z-iqz'}+iT_q^*v(i\kappa)v(q)^\dagger e^{\kappa z+iqz'}\Bigr)
 \end{equation}
Furthermore, we confirm that the transmission amplitude in this energy range (\ref{eq:originalRT-}) is analytically continued for $E_p<m$ as
\begin{equation}
T_p\to{2\over\sqrt{E_p+V_0+m\over E_p+m}+i\sqrt{E_p+V_0-m\over m-E_p}}=-{\kappa\over q}T_q,\quad
T_p^*\to{2\over\sqrt{E_p+V_0+m\over E_p+m}-i\sqrt{E_p+V_0-m\over m-E_p}}=-{\kappa\over q}T_q^*,
\end{equation}
so that the contribution coming from region iv) is expressed as integrations along imaginary $p$-axis
\begin{equation}
\int_{-im}^0{dp\over2\pi}{m\over E_p}T_pv(-p)v(-q)^\dagger e^{ipz-iqz'}+\int_{im}^0{dp\over2\pi}{m\over E_p}T_p^*v(p)v(q)^\dagger e^{-ipz+iqz'}.
\end{equation}

Putting all contributions i)$\sim$iv) together, we understand that the left-hand side of (\ref{eq:complete}) is conveniently written as
\begin{align}
&\int_{-im\to0\to\infty}{dp\over2\pi}{m\over E_p}\bigl(T_p^*u(p)u(q)^\dagger+T_pv(-p)v(-q)^\dagger\bigr) e^{ipz-iqz'}\nonumber\\
&
+\int_{im\to0\to\infty}{dp\over2\pi}{m\over E_p}\bigl(T_pu(-p)u(-q)^\dagger+T_p^*v(p)v(q)^\dagger\bigr)e^{-ipz+iqz'}.
\label{eq:TuuTvv}
\end{align}
Observe that the exponential factors make the integrands convergent at $|p|\to\infty$ in the lower-half and upper-half planes for the first and second lines, respectively, if proper branches have been chosen for momentum $q$.
Since singularities only appear on the real and imaginary axises as branch cuts, the integrals are evaluated on the imaginary $p$-axis, from $-im$ to $-i\infty$ for the first integral and from $im$ to $i\infty$ for the second.
(Remember that here we are considering the case where $z<0$ and $z'>0$.)
We note, however, that the momentum $q$ for spinor $u$, $q=\sqrt{(E_p-V_0)^2-m^2}$, and that for spinor $v$,  $q=\sqrt{(E_p+V_0)^2-m^2}$, are different quantities, though expressed with the same symbol $q$ for notational simplicity, and we have to choose a proper phase when they are analytically continued.

Actually, if we put $p=\pm i\kappa$ ($\kappa\ge m$) and thus $E_p=\pm\sqrt{\kappa^2-m^2}\equiv\pm i\Ek$ on the imaginary $p$-axis, we have, for spinor $u$, 
\begin{equation}
q=\sqrt{(\pm i{\cal E}_\kappa-V_0)^2-m^2}=\sqrt{(\pm i\Ek-V_0+m)(\pm i\Ek-V_0-m)}\equiv q_\pm,
\end{equation}
which are different from those for spinor $v$
\begin{equation}
q=\sqrt{(\pm i\Ek+V_0)^2-m^2}=\sqrt{(\pm i\Ek+V_0+m)(\pm i\Ek+V_0-m)}=-\sqrt{(\mp i\Ek-V_0-m)(\mp i\Ek-V_0+m)}=-q_\mp.
\end{equation}
We will put together those terms that have the same $(z,z')$-dependence.
When analytically continued on the imaginary $p$-axis, the first term and the last term in (\ref{eq:TuuTvv}) have the same $(z,z')$-dependence $e^{\kappa z-iq_-z'}$.
The first term is evaluated as
\begin{equation}
\int_{-im}^{-i\infty}{d(-i\kappa)\over2\pi}{m\over-i\Ek}T_p^*u(-i\kappa)u(q_-)^\dagger e^{\kappa z-iq_-z'},
\end{equation}
while the last term as
\begin{equation}
\int_{im}^{i\infty}{d(i\kappa)\over2\pi}{m\over i\Ek}T_p^*v(i\kappa)v(-q_-)^\dagger e^{\kappa z-iq_-z'}.
\end{equation}
Observe that the amplitude $T_p^*$ is actually the same quantity for both cases, for
\begin{equation}
\hbox{\rm $T_p^*$ (for spinor $u$)}
={2\over\sqrt{-i\Ek-V_0+m\over-i\Ek+m}+\sqrt{-\Ek-V_0-m\over-i\Ek-m}}
={2\over\sqrt{i\Ek+V_0-m\over i\Ek-m}+\sqrt{\Ek+V_0+m\over i\Ek+m}}=\hbox{\rm $T_p^*$ (for spinor $v$)}.
\end{equation}
After the spin sum, the spinor parts are explicitly written down as
\begin{equation}
u(-i\kappa)u(q_-)^\dagger\equiv u(p)u(q)^\dagger\Bigr|_{p=-i\kappa}
={1\over2m}\begin{pmatrix}\sqrt{(-i\Ek+m)(-i\Ek-V_0+m)}&\sigma_z\sqrt{(-i\Ek+m)(-i\Ek-V_0-m)}\\\noalign{\smallskip}
                                                  \sigma_z\sqrt{(-i\Ek-m)(-i\Ek-V_0+m)}&\sqrt{(-i\Ek-m)(-i\Ek-V_0-m)}\end{pmatrix}
\end{equation}
and
\begin{equation}
v(i\kappa)v(-q_-)^\dagger\equiv v(p)v(q)^\dagger\Bigr|_{p=i\kappa}
={1\over2m}\begin{pmatrix}\sqrt{(i\Ek-m)(i\Ek+V_0-m)}&\sigma_z\sqrt{(i\Ek-m)(i\Ek+V_0+m)}\\\noalign{\smallskip}
                                                 \sigma_z\sqrt{(i\Ek+m)(i\Ek+V_0-m)}&\sqrt{(i\Ek+m)(i\Ek+V_0+m)}\end{pmatrix}.
\end{equation}
We carefully examine the phases of the square roots and understand that these two spinor parts have opposite signs and are canceled to each other.
This means that the first term and the last term in (\ref{eq:TuuTvv}) are canceled to give vanishing contribution.
A similar argument shows that the remaining two terms, the second and the third terms in (\ref{eq:TuuTvv}), cancel to each other and we can conclude that (\ref{eq:TuuTvv}) vanishes identically.
This completes the proof of (\ref{eq:complete}) for the case of $z<0<z'$, where its right-hand side vanishes.

To summarize, {\bf Cases 1$\sim$3} complete the proof of completeness relation (\ref{eq:complete}).

\subsection{Orthogonality between $\psi^{(E)}$ and $\phi^{(E')}$}
\label{subsec:orthogonal}
As is already stated, only non-trivial relation 
is the orthogonality relation between $\psi^{(E)}$ and $\phi^{(E')}$ (\ref{eq:psidphi}).
For definiteness, the orthogonality shall be demonstrated here only for the case of $E,E'>m$.
In this case, the inner product between $\psi^{(E)}$ and $\phi^{(E')}$ is explicitly calculated as (again a trivial factor $\delta_{s,s'}$ arising from the spinor inner product shall be suppressed)
\begin{align}
&\int_{-\infty}^\infty dz\psi^{(E)}(z)^\dagger\phi^{(E')}(z)\nonumber\\
&\propto\int_{-\infty}^0dz\bigl[u(p)e^{ipz}+R_pu(-p)e^{-ipz}\bigr]^\dagger T_{q'}u(-p')e^{-ip'z}+\int_0^\infty dz\bigl[T_pu(q)e^{iqz}\bigr]^\dagger\bigl[u(-q')e^{-iq'z}+R_{q'}u(q')e^{iq'z}\bigr]\nonumber\\
&=iT_{q'}{u(p)^\dagger u(-p')\over p+p'}-iT_{q'}R_p{u(-p)^\dagger u(-p')\over p-p'-i\epsilon}-iT_p{u(q)^\dagger u(-q')\over q+q'}-iT_pR_{q'}{u(q)^\dagger u(q')\over q-q'-i\epsilon}.
\label{eq:psidphi0}
\end{align}
The second and the last terms contain terms proportional to delta functions
\begin{equation}
T_{q'}R_pu(-p)^\dagger u(-p')\pi\delta(p-p')+T_pR_{q'}u(q)^\dagger u(q')\pi\delta(q-q')
={E_p\over m}\Bigl(T_qR_p+T_pR_q{q\over p}\Bigr)\pi\delta(p-p'),
\end{equation}
which vanishes because of the relation $T_pR_q{q\over p}=-T_qR_p$.
The remaining parts are collected to yield
\begin{align}
&{iT_{q'}\over p^2-p'^2}{\sqrt{(E_p+m)(E_{p'}+m)}\over2m}\Bigl\{p-p'-(p+p')R_p-(\bigl(p-p'+(p+p')R_p\bigr){pp'\over(E_p+m)(E_{p'}+m)}\Bigr\}\nonumber\\
&-{iT_p\over q^2-q'^2}{\sqrt{(E_q+m)(E_{q'}+m)}\over2m}\Bigl\{q-q'+(q+q')R_{q'}-\bigl(q-q'-(q+q')R_{q'}\bigr){qq'\over(E_q+m)(E_{q'}+m)}\Bigr\}.
\label{eq:principalparts}
\end{align}
We note that $p^2-p'^2=E_p^2-E_{p'}^2$ and $q^2-q'^2=E_q^2-E_{q'}^2$ and furthermore
\begin{align}
p-p'-(p+p')R_p&=p(1-R_p)-p'(1+R_p)=\Bigl(\sqrt{E_q-m\over E_p-m}p-\sqrt{E_q+m\over E_p+m}p'\Bigr)T_p,\nonumber\\
p-p'+(p+p')R_p&=\Bigl(\sqrt{E_q+m\over E_p+m}p-\sqrt{E_q-m\over E_p-m}p'\Bigr)T_p,\nonumber\\
q-q'+(q+q')R_{q'}&=\Bigl(\sqrt{E_{p'}+m\over E_{q'}+m}q-\sqrt{E_{p'}-m\over E_{q'}-m}q'\Bigr)T_{q'},\nonumber\\
q-q'-(q+q')R_{q'}&=\Bigl(\sqrt{E_{p'}-m\over E_{q'}-m}q-\sqrt{E_{p'}+m\over E_{q'}+m}q'\Bigr)T_{q'}.
\end{align}
Substitution of these expressions into (\ref{eq:principalparts}) yields, after a straightforward calculation, 
\begin{align}
&i{T_pT_{q'}\over E_p^2-E_{p'}^2}{E_p+E_{p'}\over2m}\Bigl(\sqrt{(E_q-m)(E_{p'}+m)}-\sqrt{(E_q+m)(E_{p'}-m)}\Bigr)\nonumber\\
&\quad
-i{T_pT_{q'}\over E_q^2-E_{q'}^2}{E_q+E_{q'}\over2m}\Bigl(\sqrt{(E_{p'}+m)(E_q-m)}-\sqrt{(E_q+m)(E_{p'}-m)}\Bigr)\nonumber\\
&=i{T_pT_{q'}\over2m}\Bigl(\sqrt{(E_q-m)(E_{p'}+m)}-\sqrt{(E_q+m)(E_{p'}-m)}\Bigr)\Bigl({1\over E_p-E_{p'}}-{1\over E_q-E_{q'}}\Bigr),
\end{align}
which is zero because $E_q=E_p-V_0$ and $E_{q'}=E_{p'}-V_0$.
Thus the right-hand side of (\ref{eq:psidphi0}) is shown to vanish, which means that the wave function for left-incident case $\psi^{(E)}$ and that for right-incident case $\phi^{(E')}$ are orthogonal when $E,E'>m$.
The above argument would easily be extended to other energy ranges to show the orthogonality between $\psi^{(E)}$ and $\phi^{(E')}$ $\forall E,E'$.

\section{Summary and prospect}
\label{sec:summary}
The scattering states of the Dirac Hamiltonian describing fermion's dynamics under the step potential are shown to form a complete set by directly evaluating the momentum integrals.
This indirectly justifies that they are properly normalized and they constitute an orthonormal set.
Though it has been expected from a physical ground, to show that the scattering states, that is, the eigenstates of a Hamiltonian belonging to continuous eigenvalues form a complete set (when no bound states are allowed) is not at all trivial and actually has required careful analysis on their analytic properties, as expounded here in detail.
It may be stressed that in this relativistic case, due attention has to be payed also on the spinors, which has made this issue more involved than in the nonrelativistic cases.

We observe that irrelevant terms, that do not contribute to the delta function $\delta(z-z')$ and thus have to be canceled in the completeness relation, are concisely expressed as (\ref{eq:RuuRvv}), (\ref{eq:RuuRvv2}) or (\ref{eq:TuuTvv}), which would be a reflection of the existence of a more formal treatment in the relativistic case \cite{RuijsenaarsBongaarts1977}, like \cite{PalmaPradoReyes2010} in the nonrelativistic case.
The proof of the asymptotic completeness in \cite{RuijsenaarsBongaarts1977} is based on the fact that the scattering states of a particular range of energies (eigenvalues) constitute a projection operator on that energy range, which is derived from the construction of the resolvent of the Hamiltonian in terms of the scattering states.
The treatment is mathematically rigorous and elegant, however, it might seem a bit technical for physically intuitive eyes.
The presentation here, though rather involved and not elegant, would be just what such people is looking for. 
Actually, just as in \cite{TrottTrottSchnittler1989} for the non-relativistic cases, the integrations over momentum have been carried out straightforwardly and explicitly, resulting in the delta function that represents the completeness relation (\ref{eq:complete}).
It would be interesting and instructive to see how such an intuitive approach does work even for this relativistic case. 

Needless to say, the present issue is closely related to the so-called Klein tunneling (or Klein paradox) \cite{DombeyCalogeracos1999}.
In this respect, it is worth stressing that the oscillating solution of the Dirac equation in the potential region $z>0$ when the left-incident energy is below $V_0-m$ is given by a ``negative," relative to $V_0$, frequency solution $\propto Tv(q)e^{-iqz}$, see (\ref{eq:Klein}).
This is the right solution that belongs to the correct eigenvalue $E=-E_q+V_0\le V_0-m$ and satisfies the correct boundary condition, i.e., describing a positive current $j_z=\psi^\dagger\alpha_z\psi={q\over m}|T|^2\ge0$ corresponding to a transmitted wave.
The correctness would also be justified by the fact that it is one of the correct elements of the complete set. 
Other choices would result in the violation of the very eigenvalue problem and/or the boundary condition.
Therefore as already stated in \ref{sec:ScatteringStates}, there is no anomaly in the conservation of probability even in this case, but the transmission probability remains nonvanishing in the infinite potential limit $V_0\to\infty$ (\ref{eq:KleinTunneling}), which is  known as the Klein tunneling.

The complete orthonormal set obtained in this paper can be used as a basis for further explorations of such a system of fermion under the step potential within the framework of quantum field theory.
Work in such a direction is in progress and will be reported elsewhere.

\section{Acknowledgements}
H.\ N. acknowledges fruitful and encouraging discussions with Saverio Pascasio and Paolo Facchi.

\end{document}